\documentclass[twocolumn,pra,tighten,floatfix]{revtex4}

\usepackage{graphicx}
\usepackage{latexsym}

\def\<{\langle}
\def\>{\rangle}

\def\be{\begin{equation}}
\def\ee{\end{equation}}
\def\bea{\begin{eqnarray}}
\def\eea{\end{eqnarray}}

\begin{document}
\preprint{cond-mat} \title{The Eigenstate Thermalization Hypothesis in a Quantum Point Contact Geometry}

\author{G. C. Levine and B. A. Friedman}

\address{Department of Physics, Hofstra University, Hempstead, NY 11549} 

\address{Department of Physics, Sam Houston State University, Huntsville TX 77341}


\date{\today}

\begin{abstract}
It is known that the long-range quantum entanglement exhibited in free fermion systems is sufficient to "thermalize" a small subsystem in that the subsystem reduced density matrix computed from a typical excited eigenstate of the combined system is approximately thermal. Remarkably, fermions without any interactions are thus thought to satisfy the Eigenstate Thermalization Hypothesis (ETH). We explore this hypothesis when the fermion subsystem is only minimally coupled to a quantum reservoir (in the form of another fermion system) through a quantum point contact (QPC). The entanglement entropy of two 2-d free fermion systems connected by one or more QPCs is examined at finite energy and in the ground state. When the combined system is in a typical excited state, it is shown that the entanglement entropy of a subsystem connected by a small number of QPCs is sub-extensive, scaling as the linear size of the subsystem ($L_A$). For sufficiently low energies ($E$) and small subsystems, it is demonstrated numerically that the entanglement entropy $S_A \sim L_A E$, what one would expect for the thermodynamics of a one-dimensional system.  In this limit, we suggest that the entropy carried by each additional QPC is quantized using the one-dimensional finite size/temperature conformal scaling: $\Delta S_A = \alpha \log{(1/E)\sinh{(L_AE)}}$. The sub-extensive entropy in the case of a small number of QPCs should be contrasted with the expectation for both classical, ergodic systems and quantum chaotic systems wherein a restricted geometry might affect the equilibrium relaxation times, but not the equilibrium properties themselves, such as extensive entropy and heat capacity. 

\end{abstract}


\maketitle
\section{Introduction} 


Within a typical excited state of a system of noninteracting fermions, it has been shown that a sufficiently small subsystem has a density matrix that is approximately thermal, with an effective temperature proportional to the excitation energy  \cite{2015PhRvB..91h1110L}.  This is consistent with the Eigenstate Thermalization Hypothesis \cite{1991PhRvA..43.2046D, 1994PhRvE..50..888S, 2018RPPh...81h2001D, 2016AdPhy..65..239D,2018PhRvE..97a2140D, Magan:2015yoa} with entanglement being a sufficient condition in systems without interactions.  In classical statistical mechanics, if we consider a large system in the micro-canonical ensemble, the geometry of the boundary delineating a sub-system is not expected to affect the extensive properties of the equilibrium state of the subsystem; specifically equilibrium is achieved independent of how the boundary between subsystem and system is drawn and its dimensionality. If the system is ergodic, the specific boundary geometry would at most affect the equilibration time or the relaxation time if the subsystem is excited (and sub-leading terms in the free energy). It seems that similar conclusions would apply to quantum statistical mechanics insofar as the condition of ergodicity and mixing is replaced by the sufficient condition of quantum chaos. It is natural to ask, then, how geometry plays a role in equilibrium state of a degenerate fermion system where, at zero temperature, constrictions such as QPCs modify the entanglement entropy area laws and therefore possibly constrict the "equilibration" mechanism.

\begin{figure}[ht]
\includegraphics[width=8.5cm]{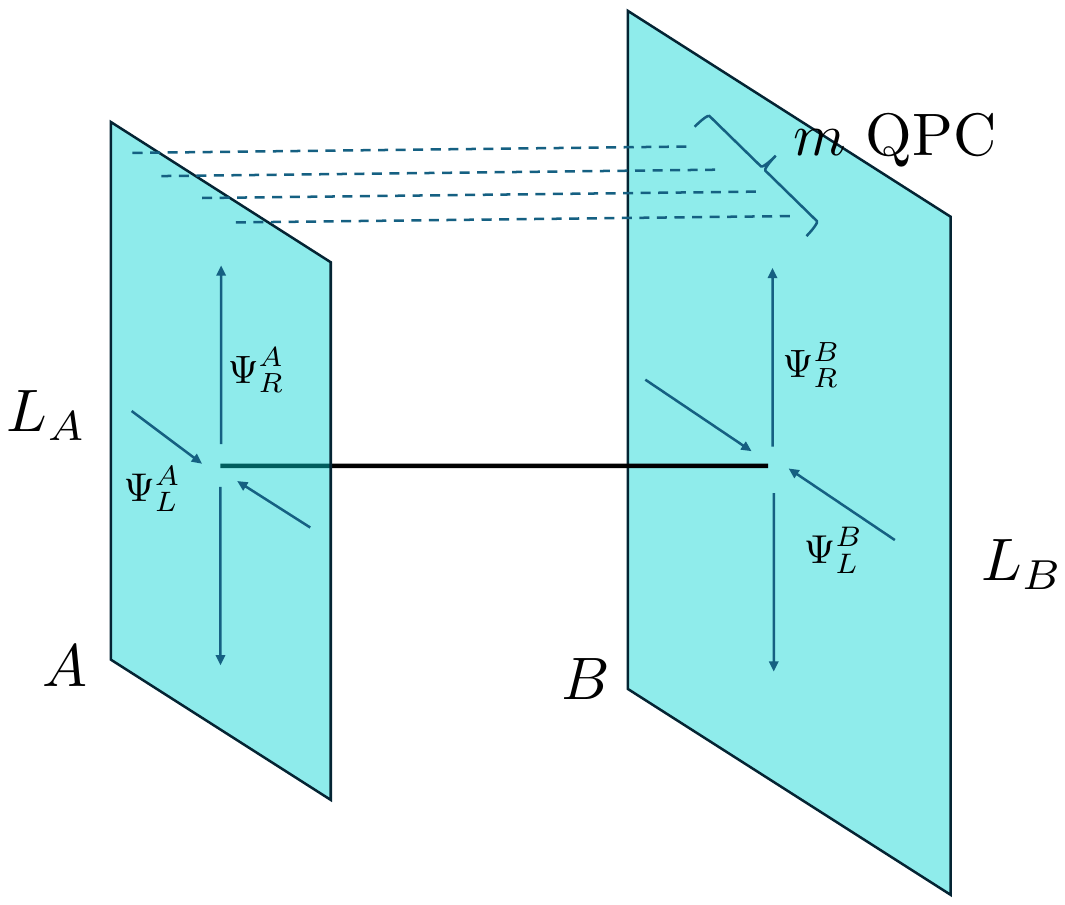}
\caption{\label{fig0}  Coupled fermion lattices $(A,B)$ of linear size $(L_A,L_B)$.  Quantum Point Contact (QPC) in center illustrates mapping to 1-d chiral fermion system. Dashed lines show a set of $m$ QPCs. If the spatial separation of QPCs is larger than $k_{\rm F}^{-1}$ they are expected to behave independently and contribute $m$ chiral channels to the entropy.}
\end{figure}

To introduce this topic generally, consider the following 2-d hamiltonian for free, non-relativistic, spin-less fermions on two periodic lattices of linear size $L_A$ and $L_B$, connected quantum point contacts (links) at sites specified by the set $\{l_{AB} \}$:

\be
\label{AB_lat_H}
H =  \sum_{i,j,\alpha }{t_{ij}^\alpha(c^\dagger_i c_j + c^\dagger_j c_i  )} 
\ee
In this hamiltonian $t^\alpha_{i,j}$ represents intra- and inter-lattice couplings: $t^A_{i,j} = -t$ where $\langle i,j \rangle \in A$; $t^B_{i,j} = -t$ where $\langle i,j \rangle \in B$; $t^{AB}_{i,j} = -t$ where $\langle i,j \rangle \in l_{AB}$. The notation $\langle i,j \rangle $ refers to nearest neighbor coupling on a square lattice (see figure \ref{fig0}). Entanglement entropy in the ground state of this model was first studied in references \cite{2008PhRvB..77t5119L, 2014PhRvA..89e2305C}.

Let us first consider the case of the ground state entanglement entropy between lattice $A$ and $B$ connected by one QPC.  A single QPC may be modeled as an s-wave scattering defect---the ingoing (outgoing) mode on the $A$ lattice is coupled to the outgoing (ingoing) mode on the $B$ lattice. Thus the entanglement entropy is effectively that of a 1-d fermionic system with size $L_A$ the linear dimension of the 2-d lattice: $S_A = b \log{L_A}$.  

\begin{figure}[ht]
\includegraphics[width=8.5cm]{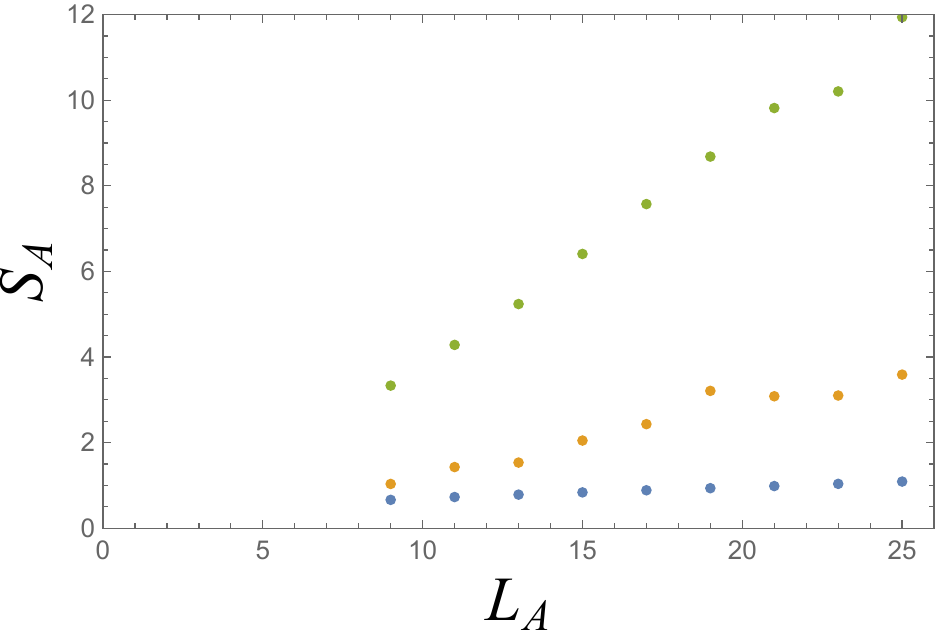}
\caption{\label{figA} Lattice size dependence ($L_A$) of entanglement entropy for a one QPC system in the ground state $E=0$ and at finite energies $E=0.1$ and $E=0.5$ (from bottom to top). Entropy for finite energy states shows proportionality to $L_A$; for comparison, entropy in the ground state is proportional to $\log{L_A}$ (see references \cite{2008PhRvB..77t5119L, 2014PhRvA..89e2305C}). For the excited state configurations, 30 "typical" excited states of the $AB$ lattice system were generated by Metropolis Monte-Carlo drawn from ensembles at temperature $T=E$ and their entropies were averaged. We will subsequently refer to the nonzero energy generated this way as an "MMC energy." (See the Numerical Methods section for further details.) The linear size of the bath lattice is $L_B=41$. } 
\end{figure}

Referring to figure 1, the $s$-wave in/out fermion modes (denoted $R$ and $L$) may be combined into a single spinor
\begin{displaymath}
\Psi^\alpha(x) = \left\{ \begin{array}{cc}
\psi_R^\alpha(x) & x>0 \\
-\psi_L^\alpha(-x) & x<0
\end{array} \right.
\end{displaymath}
where $\alpha$ refers to subsystem $A$ or $B$. Since the fermion spinor now obeys periodic boundary conditions, $\Psi^\alpha(L_\alpha) = \Psi^\alpha(-L_\alpha)$, the model may be bosonized in the standard way,
\begin{equation}
\Psi^\alpha(x) \sim e^{i\sqrt{4\pi}\phi^\alpha(x)}
\end{equation}
where $\phi^\alpha(x)$ is a right-moving boson field on the whole interval $[-L_\alpha,L_\alpha]$. The hamiltonian written in chiral form is now
\begin{eqnarray}
\label{action}
H &=& \int_{-L}^L{[ (\partial_x \phi^A(x))^2 + (\partial_x \phi^B(x))^2]dx}\\
\nonumber &-& y \cos{\sqrt{4\pi}(\phi^A(0)-\phi^B(0))} 
\end{eqnarray}
This model is equivalent to the model studied analytically at weak coupling in reference \cite{2004PhRvL..93z6402L} and numerically for arbitrary coupling in references \cite{2005JPhA...38.4327P, 2006PhRvB..73b4417Z}.  The latter studies shows that when an impurity is introduced into a 1-d system of gapless fermions, the entanglement entropy of a subsystem of length $L$, in which the impurity lies on the boundary, is proportional to $\log{L}$ with a non-universal prefactor  depending upon the impurity coupling. From the above reduction to a 1-d problem, it is expected that each sufficiently separated QPC connecting the two 2-d subsystems would contribute an entropy proportional to $\log{L}$. The entanglement entropy might then be expressed as
\be
\label{qpc_gen}
S_A = aL_A + b m \log{L_A}
\ee 
where $m$ is the number of QPCs and $b$ is a constant that depends upon the QPC coupling constant, $t^{AB}_{i,j}$.  The additive entropy $a L_A$ reflects the possible ground state degeneracy that may be present at fermion lattice occupancy $n$ close to $1/2$ for even integer $L_A$. This picture was previously confirmed numerically for the ground state of 2-d lattices coupled by a variable number of QPCs \cite{2014PhRvA..89e2305C}. (The ground state entropy for a single QPC is repeated here in figure \ref{figA} for comparison to excited state entropies). Figure \ref{figB} show the ground state entanglement entropy for three lattice sizes and a variable number of QPCs. Because the lattices are periodic with odd $L_A$ and $L_B$, and configured with a fermion density of $n=1/2$, there is no ground state degeneracy on either lattice and, therefore, no degeneracy entropy proportional to $L_A$. Each subsequent QPC contributes an entropy proportional to $\log{L_A}$ characteristic of the "radial" 1-d system described above. 

Thus, at least at zero temperature, the anomalous fermion area law may be built by constructing a 1-d boundary from multiple QPCs placed at a spacing comparable to $k_{\rm F}^{-1}$, an effective density of QPCs at which the entropy saturates.

\begin{figure}[ht]
\includegraphics[width=8.5cm]{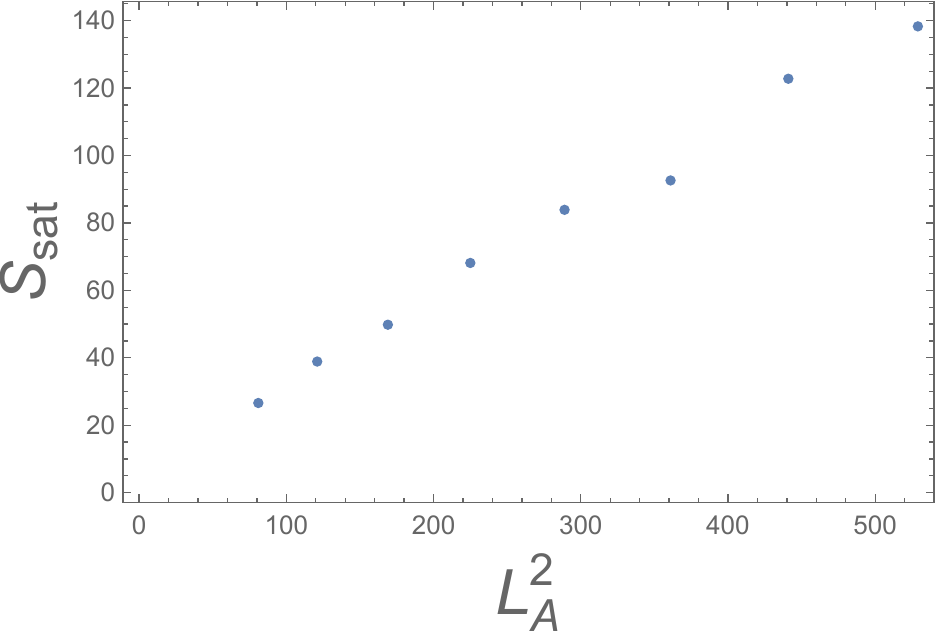}
\caption{\label{figJ} Linear lattice size dependence ($L_A$) of entanglement entropy for a QPC system with $L_A$ QPCs forming a complete 1-d boundary. The temperature $T=10.0$ chosen for the MMC energy generation is large enough compared to the bandwidth to saturate the entropy. The bath lattice $L_B=41$.  $S_{\rm sat} =S_0+ L_A^2 E_{\rm sat}$ where $E_{\rm sat} \approx 0.249$ and $S_0 \approx 8.55$. If subsystem $A$ were maximally entangled, $E_{\rm sat}$ would approach $E_{\rm sat}= \log{2} \approx 0.69$ }
\end{figure}

Alternatively, one could consider similar arguments in momentum space. Consider the subsystem $L_A$ to be imbedded within $L_B$ separated by a spatial boundary.  Following the arguments proposed successively in references \cite{2006PhRvL..96j0503G, 2010PhRvL.105e0502S, 2019arXiv190507760M} a piece of the spatial boundary of linear size $\Delta L$ also represents the one-dimensional density of states of transverse modes intersecting the boundary, $\Delta L/2\pi$. Each of the modes intersecting one piece of the boundary is then chiral and carries an entropy of $\frac{1}{6} \log{l}$, where $l$ generically represents the spatial extent of that mode within the subregion of linear size $L_A$.  Therefore the contribution to the entropy from a patch of spatial boundary within a transverse momentum interval $\Delta k$  is $$\Delta S = \frac{\Delta L \Delta k}{2 \pi} \times \frac{1}{6} \log{l}$$which, when integrated, leads to the anomalous area law. If we consider a vanishingly small section of the spatial boundary, the integral of $\Delta S$ collapses to sum over discrete "channels" consistent with quantization of phase space into minimum size blocks, $\Delta L \Delta k =2\pi$. Similar arguments are used in the construction of transverse conductance channels illustrating the quantization of electrical conductance \cite{Landauer,ImryMeso} and thermal conductance \cite{1999PhRvB..59.4992B}.  Finally, note that in the real space picture of a single QPC, the periodic fermion fields are bosonized to a single right-moving boson field \cite{Tsvelik_book}. Therefore, each QPC in the real-space picture of the anomalous area-law carries an entropy of $\frac{1}{6} \log{l}$, consistent with the momentum space treatment.

\begin{figure}[ht]
\includegraphics[width=8.5cm]{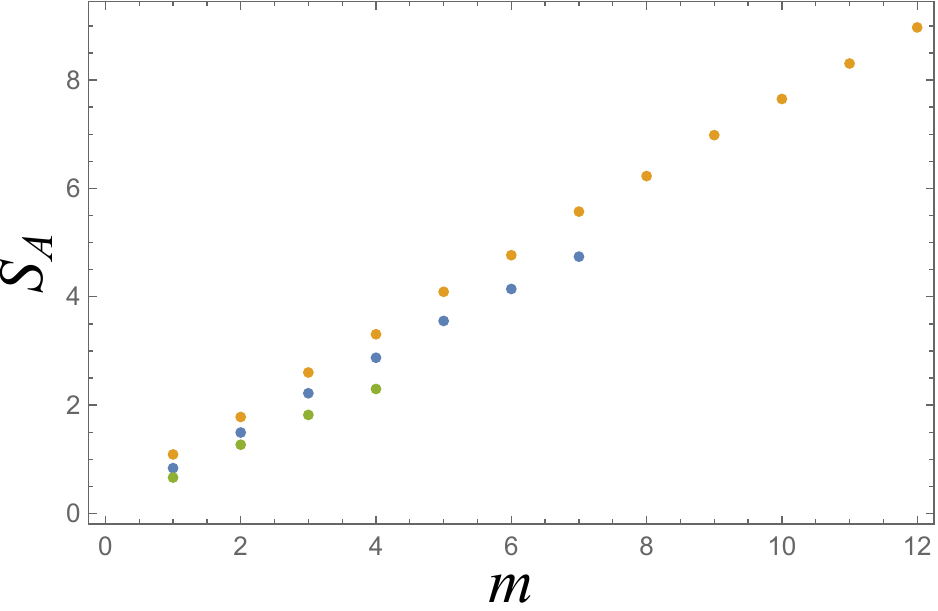}
\caption{\label{figB}  Ground state entanglement entropy dependence upon number of QPCs for 
linear lattice sizes $L_A=9,15,25$ with $L_B=41$. The slope proportional to $\log{L_A}$ consistent with equation \ref{qpc_gen} with $a=0$ and $b \approx 0.238$}
\end{figure}

\section{Results for Excited States}
Now we consider how this picture of entanglement in a QPC geometry might change for a finite energy eigenstate. The bi-partite entanglement entropy for excited states of lattice systems, including non-interacting (Gaussian) fermion systems, is now reasonably well-understood \cite{Alba:2009th, 2022PRXQ....3c0201B, 2017PhRvL.119b0601V, 2019PhRvB.100p5135J, Vidmar:2018rqk}. Generically, excited states of fermions express a volume law for the entropy and, as mentioned previously, a reduced density matrix for the subsystem that is effectively thermal \cite{2015PhRvB..91h1110L}.   

\begin{figure}[ht]
\includegraphics[width=8.5cm]{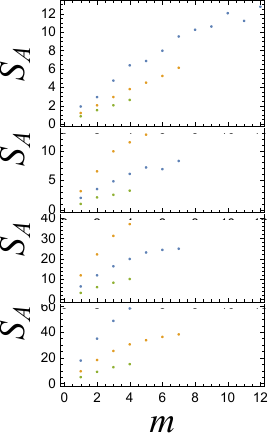}
\caption{\label{figE} Entanglement entropy of lattice $A$ ($S_A$) dependence upon number of QPCs ($m$) for linear lattice sizes $L_A=9, 15, 25 $ (from smallest to largest slope) and $L_B=41$ at average energy $E=0.05, 0.1, 0.5, 1.0$ (from top to bottom panel). Average energy specifically means "MMC Energy" $E$ as defined in the Numerical Methods section. For a small number of QPCs, the entropy is approximately additive---the graphs are at least initially proportional to the number of QPCs and the slope is proportional to $L_A$ with saturation as the number of QPCs is increased.}
\end{figure}

Now consider the coupled 2-d systems perturbatively.  At finite energy, the fermi surfaces of each lattice are now dressed with excitations in a narrow energy band around the surface. The fermi surface of each lattice then may be thought of as having quasi-degeneracies proportional to $L_{A}$ and $L_B$. These states hybridize through the perturbation of one QPC and thus produce an entanglement entropy $S_A \propto L_A$ comparable to equation (\ref{qpc_gen}) at zero temperature. We refer to this entropy as the "quasi-degeneracy" entropy---which we would expect to be evident even for one QPC---and its proportionality to $L_A$ at different energies, $E$, can be seen in figure \ref{figA}. 

As subsequent QPCs are added, it is difficult to explain an increase in entropy in terms of the quasi-degeneracy mechanism described for one QPC.  Once this quasi-degeneracy is lifted by a single QPC perturbation comparable to the band-width (and much larger than $E$), we would expect the entropy to saturate. In contrast, our calculations suggest a quasi-linear increase with QPC number. Figure \ref{figE} shows the entropy for variable number of QPCs, in different lattice sizes, and for variable excitation energy per fermion, $E$ (for details on precisely how $E$ was computed see the Numerical Methods section). In particular, figure \ref{figE}, representing the lowest energy studied ($E=0.05$), shows a behavior qualitatively very similar to that of the ground state ($E=0$, figure \ref{figB}).  

\begin{figure}[ht]
\includegraphics[width=8.5cm]{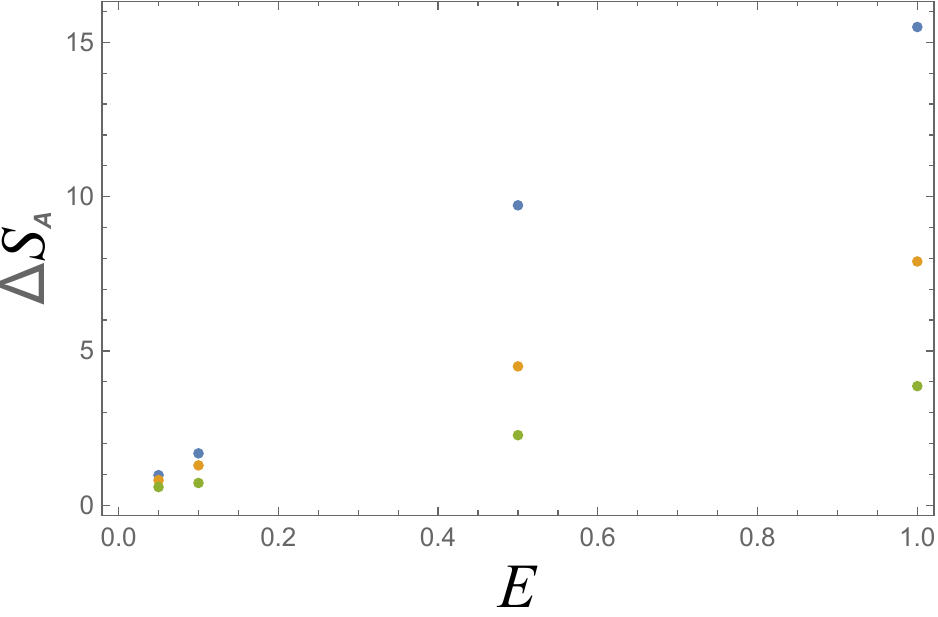}
\caption{\label{figG} Dependence of slopes computed from figure \ref{figE} upon MMC energy $E$. Slopes were computed from the first four data points for all energy ($E$) and lattice size ($L_A$) graphs using the first four data points (with the exception of $E=1.0$, in which only the first three points were used due to saturation at $m>4$.) Each additional QPC approximately contributes an additional entropy proportional to the energy $E$ of the combined $AB$ lattice system. }
\end{figure}

\begin{figure}[ht]
\includegraphics[width=8.5cm]{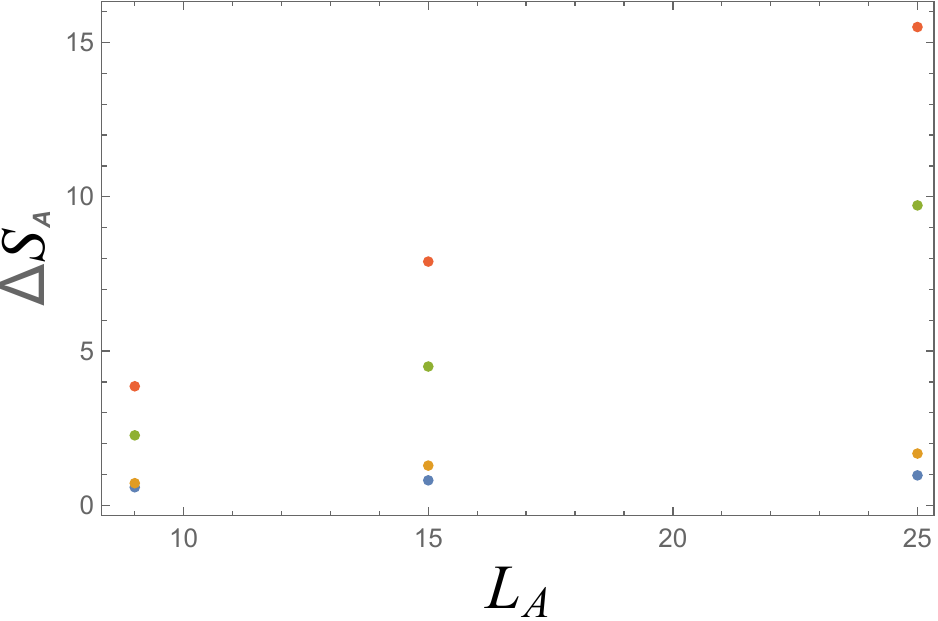}
\caption{\label{figH} Lattice $A$ linear size ($L_A$) dependence of slopes computed from figure \ref{figE}.  Each additional QPC approximately contributes an additional entropy proportional to $L_A$. Thus, the contribution of additional QPCs is seen to be sub-extensive.  }
\end{figure}

\begin{figure}[ht]
\includegraphics[width=8.5cm]{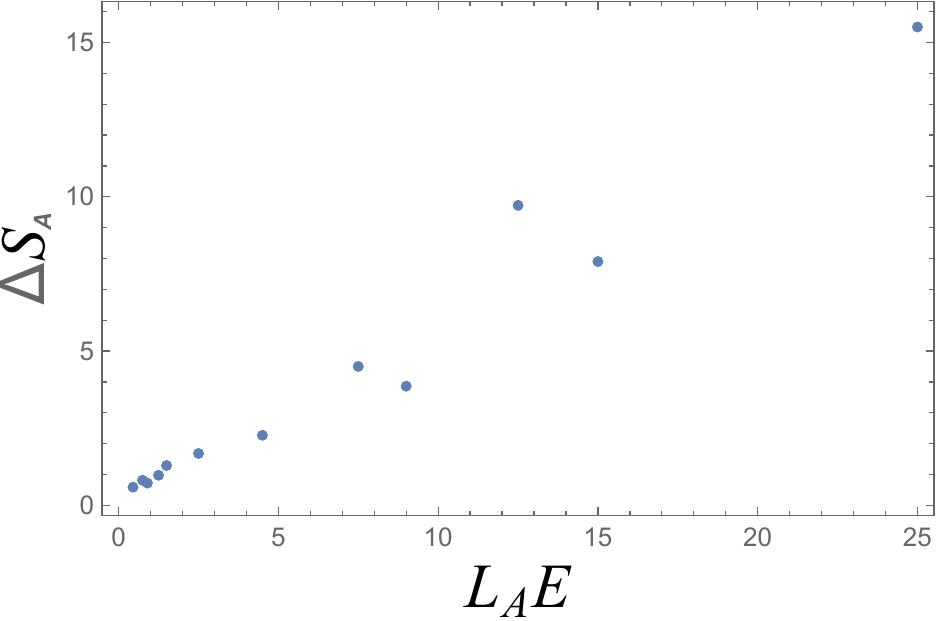}
\caption{\label{figI} Universal dependence of slopes computed from figures \ref{figE} upon the product $L_A E$.  Each additional QPC contributes an additional entropy  given approximately by $\Delta S_A = \alpha L_A E$  with $\alpha \approx 0.6$. This is the expected extensive scaling for entropy in gapless 1-d system with $E \approx T$. }
\end{figure}

The main point of this manuscript is that the entanglement entropy of finite energy eigenstates in a QPC system appear to behave the same way as their ground state counterparts: each QPC contributes a discrete additional entropy $\Delta S_A \sim L_A E$, whereas in the ground state each QPC contributes an additional entropy $\Delta S_A \sim \log{L_A}$. Thus, we argue that the additional entropy per QPC, $\Delta S_A$, follows from the entropy of a quasi-1-d gapless system (the s-wave "radial" system introduced above) at finite energy. Looking at the sequence of figures \ref{figE}, we also note that as the energy is increased, $\Delta S_A$ begins to saturate with fewer QPCs. Setting aside the saturation effect momentarily, $\Delta S_A$ may be calculated for the first several QPCs and examined for its proportionality to $E$ and $L_A$, respectively, as shown in figures \ref{figG} and \ref{figH}. In reference \cite{2019arXiv190507760M}, an insightful theory \cite{ex_state_cft} for the crossover from ground state to volume laws for the entanglement entropy was developed in which the dimensionless ratio $L_A E$ plays an analogous role to the aspect ratio of space and imaginary time, $L_A/\beta$, in a conformal field theory. Following \cite{2019arXiv190507760M}, the two scalings may be collapsed into one universal plot, figure \ref{figI}, demonstrating that the increase in entropy with each additional QPC is approximately
\be
\Delta S_A = \alpha L_A E,
\ee
the expected finite energy entanglement entropy for a gapless 1-d system.  At finite energies, we have demonstrated that equation (\ref{qpc_gen}) is modified and entanglement entropy $S_A$ approximately has the following dependence on linear lattice size $L_A$, energy, $E$, and number of QPCs, $m$
\be
\label{qpc_finite_E}
S_A = aL_A + \alpha m L_A E
\ee 
This expression appears to be valid for a small number of QPCs ($m$) or a corresponding small energy ($E$) as discussed below in the context of saturation.  An energy independent term $aL_A$ was included to accommodate the possibility of $E=0$ degeneracies as previously noted.  Comparing figure \ref{figE} at the lowest energy studied to that of the ground state in figure \ref{figB} at least visually suggest a natural crossover to E = 0 where the additive feature of entropy per QPC has been previously established. 

The sequence in figure \ref{figE} all exhibit a quasi-linear regime in $m$---most apparent for the lowest energies---where the entropy of QPCs is approximately additive. At some critical value $m_{\rm sat}$ (which seemingly depends upon energy) the entropy saturates and its increase is sub-linear. This effect may be understood by considering the saturation of entropy for a 2-d subsystem of linear size $L_A$ sharing a conventional 1-d boundary with a second 2-d lattice of linear size $L_B$. In such a case, it is well known that the entanglement entropy is extensive and proportional to eigenstate energy $E$ of the combined $L_A/L_B$ system, $S_A \propto L_A^2 E$.  This entropy is essentially equivalent to the thermodynamic entropy expected in the canonical ensemble for subsystem $A$ with temperature replacing $E$.   As the energy is increased comparable to the bandwidth, the entropy will saturate. The saturation values of entropy $S_{\rm sat}$ are also proportional to $L_A^2$, as shown in fig. \ref{figJ} and we may write:
\be
\label{S_sat}
S_{\rm sat} = S_0 + L_A^2 E_{\rm sat}
\ee
Comparing equations (\ref{qpc_finite_E}) and (\ref{S_sat}) we expect some qualitative change in behavior when $\alpha m L_A E \sim L_A^2 E_{\rm sat}$ and, consequently, we define $m_{\rm sat} = L_A E_{\rm sat}/\alpha E$.  Using parameters determined by the saturation and universal plots (figures \ref{figJ} and \ref{figI}), $m_{\rm sat} \sim 0.5 L_A/E$.  Because of this saturation effect, the slopes contributing to the universal plot were computed only from the first four data points for all energy ($E$) and lattice size ($L_A$) graphs using the first four data points (with the exception of $E=1.0$, in which only the first three points were used in that curvature from saturation was already evident at $m>4$.)

In summary, there are two potential mechanisms for entanglement entropy of the coupled lattice system, one of them reflecting the superficial entanglement of quasi-degenerate pairs of fermions on the fermi surfaces of each lattice (proportional to $L_A$), and the other, reflecting the long-range real-space entanglement of non-interacting fermions in a 1-d s-wave channel. As seen in the ground state case, studied previously, it is possible to conclude that these two entropies are essentially independent and additive, as they scale differently with $L_A$ (one linear, the other logarithmic.) In the finite energy case, these two mechanisms have the same spatial scaling $S_A \sim L_A$ and are difficult to disentangle.  However, it is hard to produce an "additive" behavior for entropy such as equation (\ref{qpc_finite_E}) from the quasi-degenerate mechanism.  And furthermore, it appears---qualitatively, at least---that the finite energy calculations (figure \ref{figE}) naturally cross over to the ground state calculation (figure \ref{figB}).   Incorporating the scaling of \cite{2019arXiv190507760M}, the most natural explanation is then that each additional QPC contributes a discrete entropy characteristic of the long-range entanglement of a 1-d gapless system at finite energy given by: 
\be
\label{S_quant}
\Delta S_A = \alpha \log{[(1/E)\sinh{(L_AE)}]}
\ee
This expression accommodates the second term of both equations (\ref{qpc_gen}) and (\ref{qpc_finite_E}).

Finally, we note that it {\sl is possible} to produce an extensive entropy by coupling the two systems with a single QPC. Returning to the quasi-degeneracy mechanism, when $L_A= L_B$, a single QPC produces an extensive entanglement entropy in an excited state $S = a L_A^2 E$ conforming to the expected entropy of the ETH.  This however appears to be a consequence of the highly symmetric (and therefore degenerate) condition brought about by $L_A= L_B$. Also, presumably, a small enough subsystem formed by creating a 2-dimensional region within $L_A$ (with a 1-dimensional boundary) would also exhibit an extensive entropy, saturating at the value of the quasi 1-d entropy of system $L_A$ as the size of the subsystem is increased.  

\section{numerical methods}

The hamiltonian (\ref{AB_lat_H}) was studied imposing periodic boundary conditions on both lattices. The terms of the hamiltonian representing the QPCs joining the lattices has coupling constants $t^{AB}_{i,j} = -t$  where $\langle i,j \rangle \in l_{AB}$  and $l_{AB}$ denotes the set of lattice sites joined between $A$ and $B$ lattices. The QPCs joined sites along diagonals in both lattices at an interval of two lattice spacings. For instance, denoting the cartesian coordinates joined by one QPC on the two lattices, $(x_A,y_A)$ and $(x_B,y_B)$, the next sequential QPC would join coordinates  $(x_A+2,y_A+2)$ and $(x_B+2,y_B+2)$.  The calculations of entanglement entropy were carried out on finite energy eigenstates $E_{\rm ex}$ of (\ref{AB_lat_H}) where the energy $E$ appearing in the calculation descriptions refers to the excited state energy per particle and is given by $E = (E_{\rm ex} - E_0)/N$ where is the many-fermion ground state energy. To generate "typical" excited states at a desired energy $E$, we adopted a Metropolis Monte-Carlo procedure, sampling at an equilibrium temperature of $E$.  Typically, we would allow 4000 MMC steps to reach equilibrium and 4000 steps between generations of subsequent eigenstates $E_{\rm ex}$.  Calculations of entanglement entropy (following procedures in \cite{2003JPhA...36L.205P}) were made for 30 eigenstates generated in this fashion and averaged; we refer to the average energy $E$ following this procedure as the "MMC energy".  Although there is a temperature involved, we emphasize that the entropies are entanglement entropies not thermodynamic entropies accorded the canonical ensemble at this temperature. The range of possible low energy calculations was limited by lattice size making it difficult to explore the crossover to ground state behavior.

\section{discussion and conclusion}

We have studied the entanglement entropy of excited states of non-interacting fermions in a two dimensional lattice system coupled by one or more QPCs. The entropy for a small 2-d subsystem ($L_A$) coupled to a larger 2-d system ($L_B$) by a small number of QPCs appears to have quasi one-dimensional characteristics---specifically, an entropy that is either logarithmic or linear in the linear dimension of the subsystem, $L_A$.  We conjecture that the increase in entropy with the addition of one QPC is given by equation (\ref{S_quant}) which interpolates between the latter two scaling behaviors based upon the dimensionless factor $L_A E$.  The only way to firmly establish the additive behavior for the entropy at finite energy, crossing over to the ground state, would be by extending the universal curve to the $L_AE << 1$ regime which we found to be computationally very difficult. 

One might consider how this result relates to the prevailing understanding of the quantum equilibration mechanism after a quench \cite{2008Natur.452..854R}. In this scenario, the many body eigenstates responsible for the superposition describing the initial state each---individually---possess characteristics of a thermal state with respect to local operators.  In the present case, the "thermal" characteristics of individual eigenstates come about from entanglement of free, noninteracting, particles \cite{2015PhRvB..91h1110L}.  Other than that, one would expect the scenario described in \cite{2008Natur.452..854R} to hold, at least for local operators. In the present case, we study an inherently non-local quantity---the entropy itself---and show that it is non-extensive, contrary to the expectation of thermal equilibrium.  The study \cite{2008Natur.452..854R} was conducted for interacting, hard core, bosons; one might question whether the presence of interactions in the fermionic case would restore the expected extensive property of the entropy. However, for a $d>1$ interacting fermi system, interactions other than in the BCS channel are irrelevant and the low energy properties are essentially identical to non-interacting fermions. It seems unlikely that the present calculation carried out with interacting fermions would produce a different dimensional dependence for the entropy.  As the present QPC geometry entropy calculation is a purely equilibrium calculation, it relies upon a heuristic understanding of how the ETH relates to dynamical equilibration.  To this end, it would interesting to study the time evolution of entropy in a non-equilibrium, quench calculation along the lines of \cite{2009PhRvA..80e3607R,2011PhRvL.107d0601S, 2024PhRvB.109l5122Y, 2024PhRvB.109v4310G}. 

Lastly, we comment on the (ir)relevance of our result to any realistic experimental scenario. Reference \cite{2013PhRvE..87d2135D} discusses entropy and heat capacity in their relation to quantities that might be calculated for a closed system and concludes that the von Neumann entropy is indeed the relevant thermodynamic entropy (see also \cite{2018PhRvX...8b1026G}). Even though the effective 1-d entropy should hold in even a 3-d geometry connected by QPCs, it is hard to imagine a realistic setting where the equilibration mechanism for fermions would be supplied exclusively by entanglement with other fermions.  In any realistic setting (e.g. electrons in a solid or effective fermions in cold atom gas), the lattices coupled by QPCs that we have described here would have large number of alternative parallel channels of entanglement through their vibrational (phonon) and radiative (photon) degrees of freedom.  von Neumann entropy would always be extensive as expected. We suggest that non-equilibrium measures of entropy, such as counting statistics and noise, that exploit conserved fermion number might be a way to access entropy in a QPC geometry \cite{2009PhRvL.102j0502K}.


\end{document}